# Migration of a Carbon Adatom on a Charged Single-Walled Carbon Nanotube

*Longtao Han[1], Predrag Krstic[1,*], Igor Kaganovich[2], Roberto Car[3]*

[1]Institute for Advanced Computational Science and Department of Material Science and Engineering, State University of New York at Stony Brook, Stony Brook, NY 11794-5250

[2]Princeton Plasma Physics Laboratory, Princeton, NJ 08543

[3]Department of Chemistry, Princeton University, Princeton, NJ 08544

**ABSTRACT:** We find that negative charges on an armchair single-walled carbon nanotube (SWCNT) can significantly enhance the migration of a carbon adatom on the external surfaces of SWCNTs, along the direction of the tube axis. Nanotube charging results in stronger binding of adatoms to SWCNTs and consequent longer lifetimes of adatoms before desorption, which in turn increases their migration distance several orders of magnitude. These results support the hypothesis of diffusion enhanced SWCNT growth in the volume of arc plasma. This process could enhance effective carbon flux to the metal catalyst.

1. Introduction

---

[*] Corresponding author. Tel: +1(865) 603-2970, Email: predrag.krstic@stonybrook.edu

Within the plasma volume, the growth of SWCNTs from transition metal catalysts can be enhanced by the flux of carbon atoms from the plasma to the external surfaces of SWCNTs. This enhanced growth results from the combined processes of adatom adsorption on the surfaces of SWCNTs and the subsequent migration of those carbon adatoms towards the metal catalyst particles [1,2]. In plasmas, nanotubes become charged through the process of being bombarded by plasma particles, specifically electrons. Here we report the effect of surface charging on the adsorption and migration of carbon adatoms, determined by performing highly accurate, all-electron Density Functional Theory (DFT) calculations, followed by Kinetic Monte Carlo (KMC) simulations. We found that while this charging only slightly affects the barriers to adatom diffusion along nanotubes, it significantly increases their adsorption energies. This seemingly counterintuitive observation is the result of increased electron density in the region between the carbon adatoms and the carbon atoms of the SWCNT. The consequence of this added density is an improvement in the covalent coupling between adatoms and SWCNTs. In concert with relatively lower diffusion energy barriers, this enhanced coupling increases the lifetime of adatoms on the surfaces of SWCNTs, allowing for longer migration distances before desorption.

Carbon nanotubes (CNT) [3, 4] have attracted significant attention from academia and industry, due to their superior thermal, mechanical, and electrical properties [5]. For example, SWCNTs have been shown to be promising materials for use in electronic [6], energy conversion [7, 8], and energy storage [9, 10] applications, because changes in tube chirality yield changes in their bandgap, and therefore allows for control over their electrical properties. The great demand for such materials calls for the development of methods for the large-scale production of minimally-defected, structurally-controlled SWCNTs.

Unfortunately the high level of control over SWCNT synthesis required for such applications has not yet been fully realized, and cannot be realized without acquiring a better understanding of the underlying growth mechanism for SWCNTs. Various models for growth mechanisms have been developed, such as the vapor–liquid–solid (VLS) [11, 12], scooter-growth [13], root-growth [14] and vapor–solid–solid models [15]. Most of the aforementioned models for the CNT nucleation and growth have been developed having in mind the chemical vapor deposition (CVD) method. In the VLS model, a carbon precursor from the gas phase adsorbs onto the surface of a transition metal catalyst particle and then dissociates. The resulting carbon atoms then diffuse into the liquid catalyst nanoparticle, possibly forming metal carbides. This process eventually results in saturation of the nanoparticles with carbon, excess of which then crystallizes at the catalyst surface. These excess carbon atoms assemble into a graphene cap whose edges are strongly chemisorbed to the metal catalyst. A crucial feature of this mechanism is its avoidance, at all stages of growth, of any open graphene edges, the existence of which would mean exposing energetically expensive dangling bonds. The energy minimized, curved surface of the cap has $sp^2$ covalently bonded carbons, forming hexagons and pentagons, with minimal dangling bonds at the edges. Once the cap is formed, insertion of new carbons occurs between the tube edge and the catalytic particle resulting in the growth of a SWCNT. Once overcoating of the catalyst by carbon is reached, the process is suppressed and deactivated [11, 12]. A proposed surface-mediated growth model can also explain the lower activation energy associated with plasma-enhanced CVD growth [16]. In this model, carbon transport toward the root of SWCNTs occurs not by bulk diffusion through a liquid particle, but rather by surface diffusion over a potentially solid catalyst particle.

High-pressure, arc-discharge plasma, utilizing carbon electrodes can be used for producing high quality, defect-free SWCNT [17]. However, despite the successful development of this method, detailed understanding of all physical and chemical processes taking place during the SWCNT synthesis in the arc is challenging. In contrast to the CVD method which has carbon precursor molecules as its only reactive species, many reactive species exist in the plasma volume, including neutral and excited atoms, ions, electrons, radicals, and molecules. Furthermore, unlike the CVD method, the temperature of an arc-discharge plasma varies from 10,000 K in the arc column to below 1000K at the discharge periphery, offering a very broad range of gas temperatures as well as a significantly higher flux of feedstock particles. Therefore, the mechanisms of nucleation and growth of the SWCNT in the arc plasma could be significantly different from those proposed for CVD synthesis. For instance, in contrast to CVD where carbon precursors must first adsorb to catalyst particles and then dissociate into reactive carbon, for arc-discharge plasma synthesis, reactive carbon atoms could be adsorbed on the SWCNT surface directly from plasma. The high temperature plasma may enable an accelerated surface migration of adsorbed atoms toward the junction between catalyst and root of the growing SWCNT, enhancing the total carbon flux available for SWCNT growth. In addition, in the arc-discharge method, the nanotube surface is subject to the flux of plasma ions and electrons, which are capable of charging the SWCNT. These distinctions from CVD may be of importance in arc-discharge plasma synthesis, achieving accelerated, controlled growth of SWCNT, and thus are the main motivations for the present study.

The adsorption behavior of atoms and molecules on SWCNT has been studied extensively, mainly for the purpose of hydrogen storage [18], gas sensing [19], catalysis [20] etc., but few of these studies have focused on the adsorption of carbon atoms on SWCNT. Durgun et al. [21] utilized a pseudopotential plane-wave DFT (PWDFT) method to calculate the stable adsorption geometries and adsorption energies ($E_a$) for a number of adatoms on SWCNT, ranging from alkali to transition metals and group IV elements. They found adsorption energies, $E_a$= 3.7 and 4.2 eV at the two stable positions of a C adatom on (8,0) SWCNT [20]. Lehtinen et al. [22] examined adsorption of carbon adatoms on graphene using spin-polarized PWDFT, finding the equilibrium position to be a bridge-like structure with $E_a$=1.40 eV. Krasheninnikov et al. [23, 24] calculated $E_a$ making use of the tight-binding DFT approximation (DFTB) and PWDFT to study adsorption and migration of carbon adatoms on SWCNT's of various chiralities. Their calculations for the adsorption energies of carbon atoms at their most stable sites on SWCNTs ranges from 2.7-4 eV for tubes with diameters from 0.6-1.4 nm. The migration barriers of carbon adatoms are found to be in the range of 0.6–1 eV for SWCNT with typical diameters of 1-1.4 nm, and are governed by the orientation of the C-C bond with respect to tube axis [23, 24].

Studies of the effects of charging on the properties of SWCNTs have been conducted both experimentally and theoretically, to determine the charge density profiles and relevant electronic structures. Keblinski et al. [25] conducted DFT calculations of charge distributions on negatively and positively charged nanotubes of finite length. They found that the charge distributed in U-like profile along the nanotube for both positive and negative charges, with charge density located primarily at the tube ends. Follow-up studies showed that the charge enhancement at the tube ends decreases as the tube length increases [26], which was later experimentally confirmed by showing that over a micron long nanotube the charge was almost uniformly distributed [27]. In addition, recent research performed by Wang et al. showed that CNTs can be negatively charged during thermal CVD growth, and that charging can be utilized to control their resulting chirality [28].

2.  **Computational details**

We studied the charging effects for metallic (5,5) SWCNT in the armchair configuration, which exhibits surface migration paths with the lowest energy barriers along the CNT axis. The calculations were carried for a (5,5) SWCNT of finite length (2 nm), containing 180 carbon atoms. In order to pacify dangling bonds at the CNT ends, we terminated each end with "crown" of 10 hydrogen atoms. Hydrogen-termination ensures chemical passivation of the dangling carbon bonds at the edges, which provides faster convergence of the DFT calculations [29]. This is a common approach in the studies of CNT bulk properties [30]. Varying the negative charge on the CNTs from 0 to -12e in steps of -2e, (where e is the elementary charge), we identified the stable sites, and then calculated adsorption energies, diffusion energy barriers, and vibrational frequencies. In addition to the metallic (5,5) SWCNT, we studied the migration parameters of two semiconducting SWCNTs (adsoprtion energies, and diffusion barriers) (10,5) and zig-zag (10,0), of finite lengths in the absence of charging.. These computations were performed using all-electron molecular DFT [31, 32] with 6-31G* valence double-zeta polarized Gaussian basis set and PBE0 hybrid functional [33], implemented by NWChem computational chemistry package [34]. The results are shown in the Supporting Information (SI), Fig. S1 and S2. We find that only (5,5) SWCNT has a continuous migration path with low diffusion barriers for adatom migration, in the direction of the tube axis. This is the reason to study in detail the effects of charging on adatom migration on the external surface of the metalic, armchair (5,5) SWCNT.

*2.1 Adsorption sites and energies*

Three equilibrium adsorption sites for carbon adatom were identified for a (5,5) SWCNT, close to the middle of the CNT length. In agreement with previous studies [20, 22], they are located above the bridges connecting neighboring carbon atoms, as shown in Fig. 1(a). Note that sites 1 and 3 in Fig. 1(a) are equivalent. The location of the equilibrium adsorption sites in the middle of the bridging sites is not unique to (5,5) SWCNTs; two examples showing adatom adsorption sites at corresponding bridge sites of SWCNT of other chiralities are shown in Fig. S1. The geometry of SWCNT was optimized for each charge, with and without an adatom present. The adsorption energy is calculated from

$$E_a^q = E(CNT + C)^q - E(CNT)^q - E(C) \qquad (1)$$

where superscript "q" is the charge of the system, and E(CNT+C), E(CNT) and E(C) are the geometry optimized configuration energies of CNTs with an adatom present, a pristine CNT, and spin-polarized (triplet) single carbon atom, respectively. We assume that the CNT is charged before adsorbing the adatom, as expressed by $(CNT)^q$ in Eq. 1. Previous studies [20, 21] indicated that adsorption energies are dependent on the local curvature of SWCNT, showing that adsorption energy is lowest when carbon-carbon bond orientation is along the tube axis, and is highest when the orientation is perpendicular to the tube axis, (in agreement with our findings, Table 1).

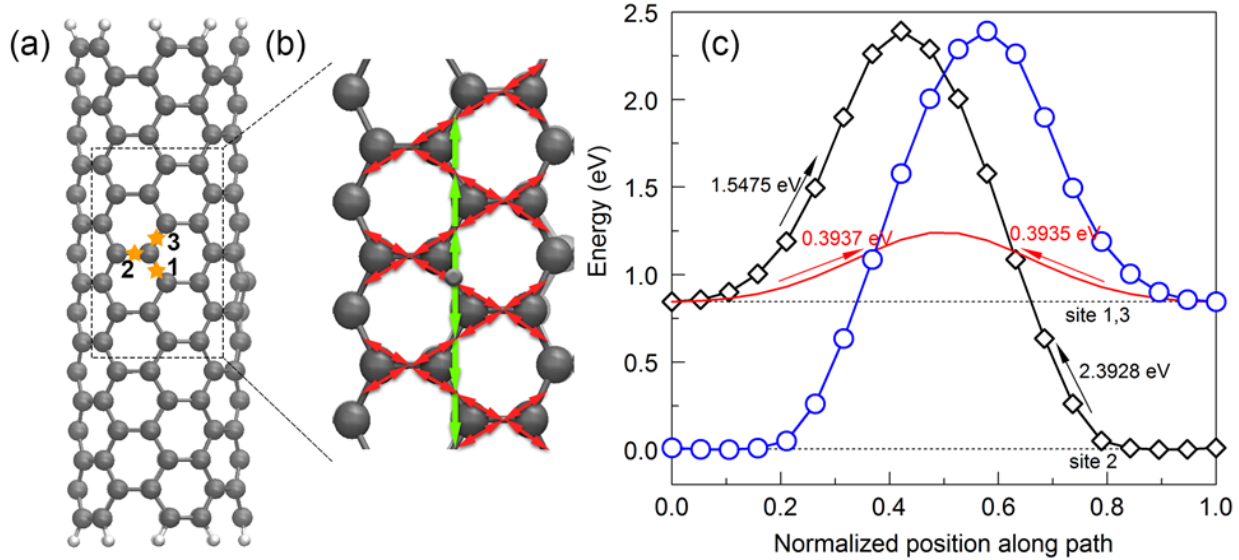

**Fig. 1.** (a) Equilibrium adsorption sites marked by yellow stars, (b) possible migration paths and (c) energy profiles of migration barriers between the equilibrium sites for an adatom on a (5,5) SWCNT. The green arrows in (b) represent the paths with the lowest energy barriers, which are between sites 1 and 3. The red arrows represent paths through sites 2 with barriers higher than 1.5 eV, and these paths are less probable routes for migration than those indicated by the green arrows. Relative energies of adatoms adsorbed at three sites are illustrated by dashed horizontal lines in (c), and the energies of an adatom along the migration paths between sites 1-2, 1-3 and 2-3 are represented by hollow rhombs on a black line, a red line, and hollow circles on a blue line, respectively.

Using Eq. 1 we also calculated adsorption energies as a function of the number of negative charges present on a (5,5) SWCNT, as shown in Fig. 2 and Table S1. For all three adsorption sites, we found adsorption energies increased with increasing charge. This increase in energy is particularly large when |q|> 4e, for reasons that will be discussed in detail in the following section.

**Table 1**

Adsorption energies at the three types of adsorption sites for SWCNTs of each of the three chirality types. Data in Refs. [22,23] are calculated by the PWDFT, with PAW potential and GGA hybrid functional. Our calculations are performed on the finite-length SWCNTs, using molecular DFT in NWChem package, with polarized Gaussian basis 6-31G* and PBE0 hybrid functional.

| Chirality type | Adsorption energy at three adatom sites (eV) | | |
|---|---|---|---|
| | Site 1 | Site 2 | Site 3 |
| (5,5) | 2.01 | 2.87 | 2.01 |
| (5,5) by PWDFT ref. 22,23 | 2.35 | 3.30 | 2.35 |
| (10,5) | 1.64 | 1.99 | 1.35 |
| (10,0) | 1.63 | 2.29 | 2.29 |
| (10,0) by PWDFT ref. 22,23 | 2.10 | 2.60 | 2.60 |

*2.2 Migration barriers*

Identification of the minimum energy path (MEP) for migration from one stable equilibrium configuration to another was done by the nudged elastic band method, and implemented by DFT in NWChem. Each energy barrier height ($E_b$) is listed as the difference between the transition state energy and the equilibrium energy (Fig. 2 and Table S2). Surface migration paths for an adatom between the equilibrium sites for the armchair (5,5) SWCNT are shown in Fig. 1(b). Migration paths with high-energy barriers between the stable sites are shown in red. Possible fast migration paths, corresponding to the low-energy barriers are shown in light green, which occur in the direction of the CNT axis. The corresponding MEPs are shown in Fig. 1(c). Even though the most stable adsorption states are those with the highest binding energy, which are regularly connected with large barriers, it is worth noting that bombardment of the CNT by plasma particles may bring adatoms to vibrationally excited states and thus may accelerate the migration. As mentioned above, there is no continuous low-barrier path for the SWCNTs of the other two considered chirality types. The path with lowest barriers for (10, 0) and (10, 5) SWCNT still contains a barrier of 1.1 eV and 0.7 eV, respectively, as shown in Fig. S2.

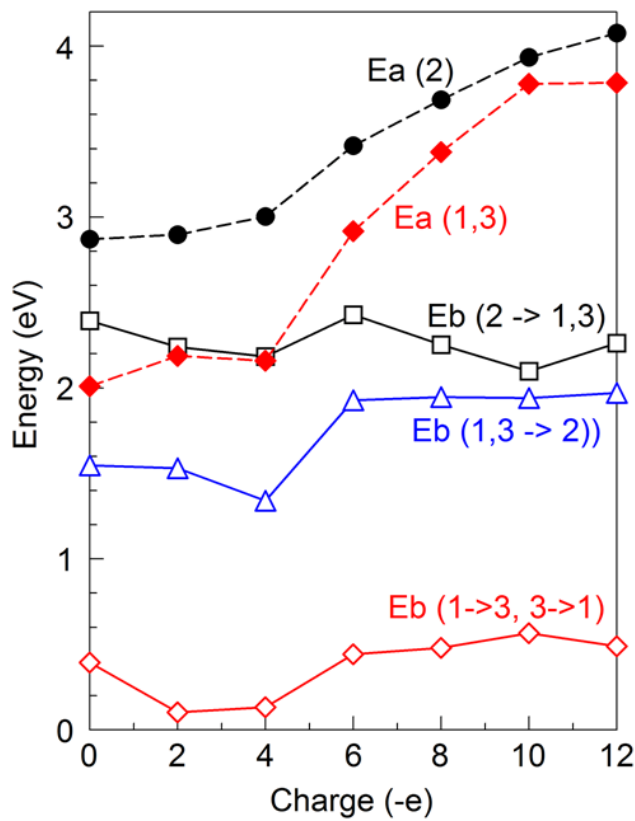

**Fig. 2**. Migration barrier heights ($E_b$) and adsorption energies ($E_a$) of a (5,5) SWCNT as a function of charge on the SWCNT.

Unlike adsorption energies, which monotonically increase by almost 90 % in the considered range of charges, the migration barriers, shown in Fig. 2 and Table S2, non-monotonically vary up to 30%. In the next section we will show simulation results for the migration of an adatom over a charged (5,5) SWCNT using the KMC method.

*2.3 Kinetic Monte Carlo*

Simulations of carbon adatom migration were conducted using the KMC method [35]. KMC simulations account for desorption from, and hopping between, the three types of adatom sites. The corresponding transition rates were calculated using the obtained energy barriers, adsorption energies, and vibrational frequencies obtained using Arrhenius-type formulas:

$$k_{s \to f} = \nu_s \exp(-\frac{E_{s \to f}}{k_B T}) \qquad (2)$$

$$k_d^s = \nu_s \exp(-\frac{E_a^s}{k_B T}) \qquad (3)$$

where $\nu_s$ is the vibrational frequency of the corresponding state in the direction of reaction, $E_{s \to f}$ is the energy barrier for the transition from a state "s" to a state "f," $E_a^s$ is the adsorption energy of a state "s," and temperature $T$ was chosen to be 1700 K.

The vibrational frequencies, $\nu_s$, drivers for both hopping and desorption, were calculated by numerical Hessian in NWChem, and the results are presented in the SI. Table S3 shows the normal mode frequencies for an adatom located at the different sites on a SWCNT with varying charges. These frequencies correspond to three modes illustrated in Figure S3. All normal mode frequencies are close to $10^{13}$ Hz, in agreement with previous studies [23]. The vibrational frequency of an adatom is a projection of the normal mode frequencies to the direction of the reaction coordinate.

Transition rates for the KMC simulation, hopping rates as determined by Eq. 2 and desorption rates as determined by Eq. 3, are tabulated in Table S4. All trajectories were started in an adsorbed state. Lifetime is defined as the total time an adatom stayed on the surface of a CNT before desorbing into the vacuum. Migration distance is defined as the distance that an adatom migrated along the tube axis direction, during its lifetime. Each migration trajectory contains up to $10^{10}$ steps. We calculated the average values from 400,000 trajectories at each charge to get the convergence. In this calculation the length of CNT was assumed infinite, with the properties as calculated in the middle of our finite tube. The charge density with the total added charge q at the tube is used as q/L (e/nm), where L=1.98 nm.

## 3   Results and discussion

Both migration distance and lifetime of the migration process are shown in Fig. 3 for a CNT with varying numbers of charges q. While we identified three possible initial sites for an adatom, in this paper we show only the case where site 2 is initially populated. Because the other cases, where sites 1 and 3 are initially populated, almost fully overlap with the case where site 2 is initially populated. This is valid, because $E_b$ (1,3->2) < $E_a$ as shown in Fig.2 and adatom transfers to the site 2 well before it can move far along CNT before desorption. The migration distance increases as q increases – from 25 nm at q=0, to about 100 nm at q=-4e, and then further increases to approximately 13 microns at q=-12e. On the other hand, the lifetime slowly increases – from 10 $\mu$s at q=0, to 27 $\mu$s at q=-4e, and finally a significant increase to about 160 ms at q=-12e. The abrupt increase of migration distance at q=-2e results from the lower migration energy barriers between sites 1 and 3 at q=-2e than at q=0 or -4e. While a slight increase in the migration energy barrier partially offsets the effect of the increased adsorption energy, the migration distance and lifetime increase significantly only for $|q| > 4e$, where there is a corresponding noticeable increase in the adsorption energy as well.

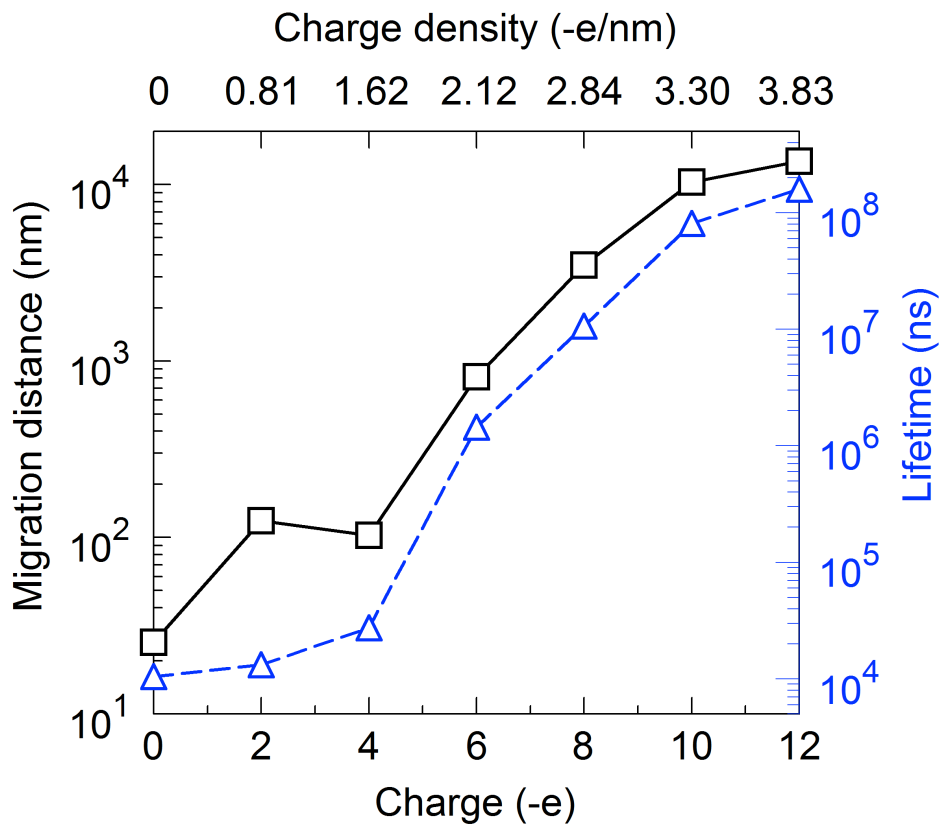

**Fig. 3.** Migration distance, (the black line with squares), and lifetime, (the blue dashed line with triangles), of a carbon adatom as a function of the charge on a (5,5) SWCNT of 2 nm long. Charge density in the middle of the tube for each case is shown on the upper axis.

As others have previously reported, for finite tubes which are not terminated by hydrogen atoms, the charge density on a charged SWCNT is primarily located at the tube ends, (a U-like profile) [25, 26]. In our case, we found that termination of SWCNTs with hydrogen atoms results in polar carbon-hydrogen bonds. These observed dipoles show a positive charge to be on the hydrogen atoms and a negative charge to be on the carbon atoms (see Fig. 4). Away from the SWCNT ends, the charge density is nearly homogeneous. These distributions of charges were determined using

Mullikan population analysis. While the absolute value of charges determined based on Mulliken analysis is not reliable, the trends in charge differences between sites for varying total SWCNT charge should be reliable. As is evident in Fig. 4 and Table S5, when an adatom is added to a SWCNT with varying charges, it changes from having a positive charge of approximately 0.26e (for q=0) to having a negative charge of approximately -0.3e (for q=-12e). For a centrally located carbon atom in a SWCNT, the negative charge localized on it does not exceed 0.05e. The corresponding repulsive electrostatic energy between an adatom and a CNT is expected to increase with increased CNT charging.

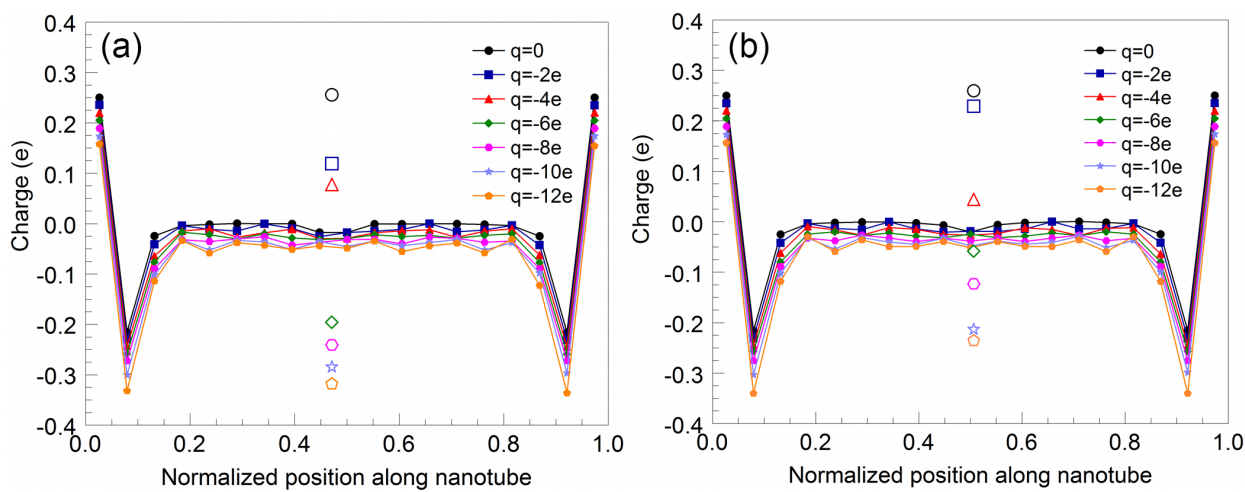

**Fig. 4.** The distribution of Mulliken charges along the axis of a (5,5) SWCNT for an adatom adsorbed at (a) site 1, (b) site 2. Filled symbols represent the charges localized on the carbon atoms of a SWCNT. Each data point represents the charge per atom averaged over 10 atoms that share the same normalized position along nanotube. Hollow symbols represent the charges localized on an adatom.

As shown in Fig. 2 and Table S1, the adsorption energies increase significantly when the negative charges on a CNT exceed 4e. This increase is associated with an increase in covalent bonding between an adatom and the two carbon atoms closest to it on the surface of a SWCNT. This is illustrated in Fig. 5, showing the differences of electron densities in the vicinity of adatom of the charged CNT (q=-4e, -8e, and -12e) + adatom, and charge-neutral cases. Although the Mulliken negative charge of the adatom increases with q, this increase is not localized on the adatom, but rather occupies the space between the adatom and two carbon atoms closest to it on CNT. This increased density strengthens the covalent bonding and thus contributes to the increased adsorption energy. This result is consistent with the decrease of the HOMO-LUMO gap of the

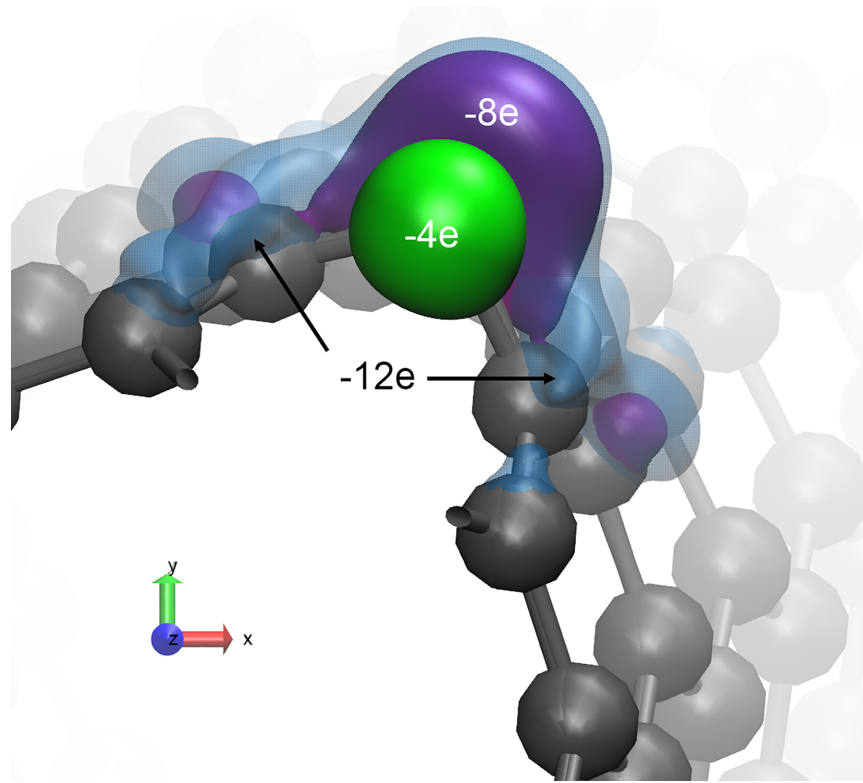

**Fig. 5.** Isosurface of the difference in electron density with additional charge: -4e (green), -8e (purple), -12e (dim blue), compared to the q=0 case. The isovalue of 0.02 is used in all three cases.

CNT + adatom with increasing q, as shown in Table 2. While Fig. 5 illustrates the electron density changes for the stable site 2, the adsorption energy for sites 1 and 3 increases even more for |q| > 4e. This occurs because the equilibrium adsorption configuration changes. As shown in Fig. 6, three carbon atoms of the SWCNT bond to the adatom instead of only two as discussed previously. Interestingly, the migration barriers do not change significantly with increasing q, (Fig. 2 and Table S2). The barrier along the fast migration path, (the green path in Fig. 1(b)), is less than 0.5 eV, even for q=-12e. Because the desorption rates are exponentially dependent on $E_a^s$ (Eq. 3), and adsorption energies increase in absolute value with q, increasing q greatly increases the lifetime of adatoms, yielding increases in the migration distances, (as obtained by KMC and shown in Fig. 3).

**Table 2**

HOMO-LUMO gap of a (5,5) SWCNT (with an adatom present) for different charges

| Molecule type | HOMO-LUMO Gap (eV) | | | | | | |
|---|---|---|---|---|---|---|---|
| | q=0 | q=-2e | q=-4e | q=-6e | q=-8e | q=-10e | q=-12e |
| CNT | 1.4170 | 0.8367 | 1.2217 | 0.8620 | 0.7700 | 0.5628 | 1.0759 |
| CNT+C(1) | 1.3780 | 0.9406 | 0.9049 | 0.9527 | 0.8001 | 0.7253 | 0.9130 |
| CNT+C(2) | 1.4869 | 0.7152 | 1.0255 | 0.8767 | 0.7455 | 0.5675 | 0.8176 |
| CNT+C(3) | 1.3769 | 0.9392 | 0.9057 | 0.9522 | 0.8002 | 0.7259 | 0.9130 |

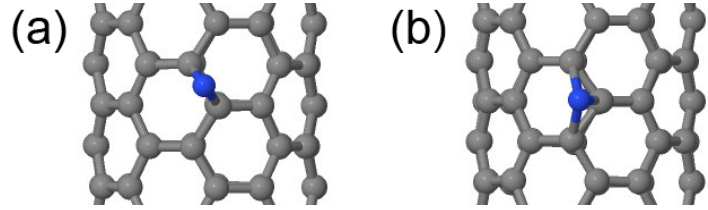

**Fig. 6.** Equilibrium adsorption configuration of an adatom at site 1 for (a) |q| ≤ 4e; (b) |q| >6e.

Based on the above results, we can estimate the contribution from surface migration to the carbon flux toward the root of a SWCNT. If homogeneous flux density of carbon particles onto a SWCNT is $F$, and $r$ is the radius of the catalyst particle, the total number of particles impinging on the hemisphere of the catalyst particle in a unit time is $2\pi r^2 F$. Assuming the radius of SWCNT is correlated to the radius of catalyst, the number of particles in unit time impinging on the SWCNT surface of length $L$ is $2\pi r L F$. The ratio of these fluxes is $L/r$. Since $L$ can be on the order of 10 microns, while $r$ is typically on order of nanometers, this ratio can reach several orders of magnitude. A significant proportion of these adsorbed carbon atoms on the SWCNT surface could diffuse toward the catalyst-CNT root. Possible reduction of this fluence at the junction of a SWCNT and catalyst particle [36, 37, 38] is beyond the scope of the present study. Healing of defects during the CNT growth can also be aided by enhanced carbon adatom surface migration. As such, it is possible that enhanced migration is responsible for the production of the characteristically low-defect SWCNT synthesized in the arc plasma volume [39].

# 4 Conclusions

In summary, we explored the adsorption and migration behavior of a carbon adatom on negatively charged, armchair SWCNT of finite length, finding significant increase in the migration distance when $|q| > 4e$. The transition rates were determined using first-principles DFT calculations. The increased adsorption energies for negatively charged SWCNTs result from increased covalent bonding between the adsorbed carbon adatom and the carbon atoms of CNTs. This stronger bonding leads to significantly increased lifetime for the adatom, allowing for longer migration distance before desorption back into the plasma. These findings indicate an enhanced carbon adatoms flux on the external surface of SWCNTs toward the metal catalyst, which could lead to a profound increase in the growth rate of SWCNTs in the arc plasma volume.

**Acknowledgments**

This work was supported by the U.S. Department of Energy, Office of Science, Basic Energy Sciences, Material Sciences and Engineering Division Grant No. DE-AC02-09CH11466. Presented results were in part calculated using XSEDE computing facilities, (Stampede and Comet), LIRed computing facilities of IACS of SBU, and DOE NCCS computing facilities at ORNL. We are grateful to Prof. Robert J. Harrison for inspiring discussions.## Appendix A. Supplementary data

Supplementary data related to this article can be found at *URL*.

# Supporting Information

# Migration of Carbon Adatom on a Charged Single Walled Carbon Nanotube

*Longtao Han[1], Predrag Krstic[1,*], Igor Kaganovich[2], Roberto Car[3]*

[1]Institute for Advanced Computational Science and Department of Material Science and Engineering, State University of New York at Stony Brook, Stony Brook, NY 11794-5250

[2]Princeton Plasma Physics Laboratory, Princeton, NJ 08543

[3]Department of Chemistry, Princeton University, Princeton, NJ 08544

---

[*] Corresponding author. Tel: +1(865) 603-2970, Email: predrag.krstic@stonybrook.edu

# S1: Migration parameters in absence of charging of two semiconducting SWCNT of (10,5) and zig-zag (10,0) chiralities with finite lengths

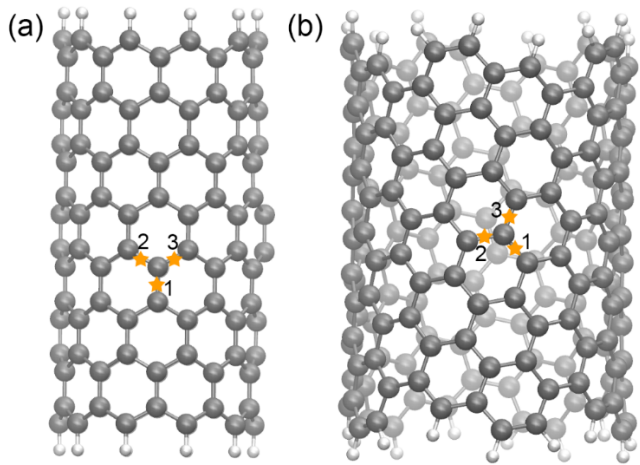

Figure S1 Equilibrium adsorption sites for SWCNT of two chirality types: (a) (10,0); (b) (10,5)

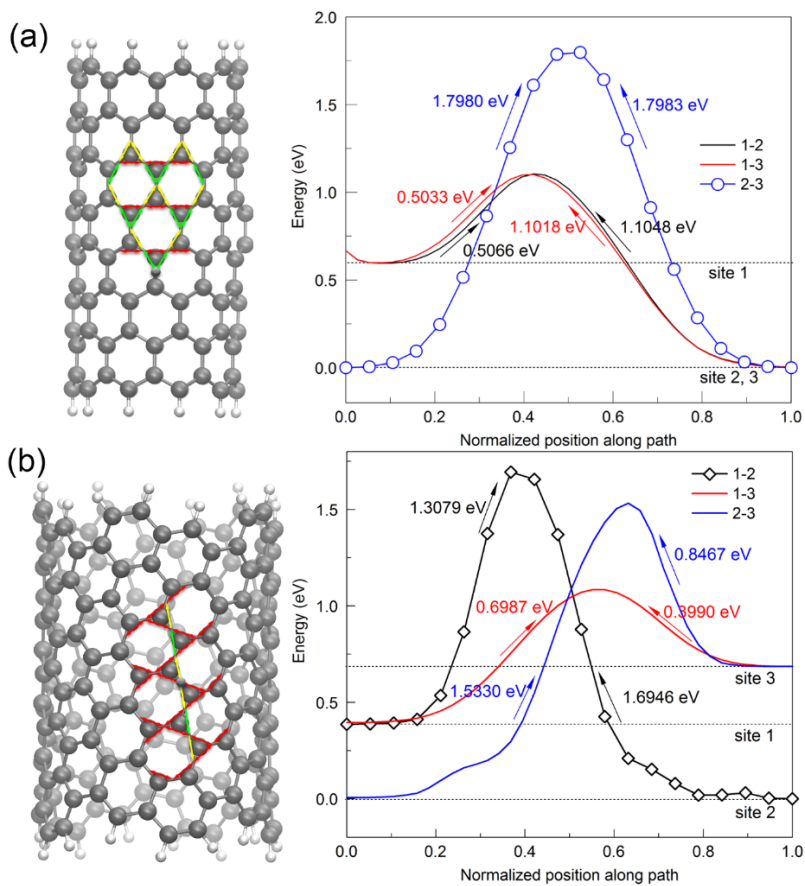

Figure S2 Possible migration path and energy profile along path of (a) (10,0) and (b) (10,5) SWCNT

## S2: Migration parameters of charged SWCNT of (5,5) chirality with finite lengths

**Table S1** Adsorption energy of adatom on (5,5) SWCNT at different charges

| Adsorption site | Adsorption energy (eV) | | | | | | |
|---|---|---|---|---|---|---|---|
| | q=0 | q=-2e | q=-4e | q=-6e | q=-8e | q=-10e | q=-12e |
| 1 | 2.0106 | 2.1876 | 2.1603 | 2.9171 | 3.3800 | 3.7796 | 3.7860 |
| 2 | 2.8715 | 2.8969 | 3.0040 | 3.4177 | 3.6873 | 3.9357 | 4.0773 |
| 3 | 2.0104 | 2.1880 | 2.1607 | 2.9172 | 3.3801 | 3.7797 | 3.7860 |

**Table S2** Migration energy barriers of adatom on (5,5) SWCNT at different charges

| Direction | Migration energy barrier (eV) | | | | | | |
|---|---|---|---|---|---|---|---|
| | q=-0 | q=-2e | q=-4e | q=-6e | q=-8e | q=-10e | q=-12e |
| 1,3->2 | 1.5476 | 1.5314 | 1.3400 | 1.9277 | 1.9462 | 1.9418 | 1.9703 |
| 2->1,3 | 2.3928 | 2.2408 | 2.1837 | 2.4283 | 2.2535 | 2.0980 | 2.2616 |
| 1->3 3->1 | 0.3935 | 0.1025 | 0.1328 | 0.4429 | 0.4793 | 0.5647 | 0.4896 |

## S3: Vibrational frequencies of adatom on (5,5) SWCNT

**Table S3** Normal mode frequencies of adatom at different charges

| Charge | Normal mode frequency ($10^{13}$ Hz) | | |
|---|---|---|---|
| | Site 1 | Site 2 | Site 3 |
| q=0 | 0.912 | 0.457 | 0.913 |
| | 1.415 | 2.218 | 1.415 |
| | 2.387 | 2.514 | 2.386 |
| q=-2e | 0.959 | 1.092 | 0.957 |
| | 1.376 | 2.252 | 1.377 |
| | 2.368 | 2.556 | 2.369 |
| q=-4e | 0.869 | 1.669 | 0.866 |
| | 1.395 | 2.260 | 1.393 |
| | 2.370 | 2.402 | 2.370 |
| q=-6e | 0.847 | 0.894 | 0.839 |
| | 1.357 | 2.214 | 1.356 |
| | 2.348 | 2.466 | 2.347 |
| q=-8e | 0.878 | 1.367 | 0.872 |
| | 1.387 | 2.293 | 1.385 |
| | 2.392 | 2.497 | 2.391 |
| q=-10e | 0.828 | 1.074 | 0.827 |
| | 1.366 | 2.205 | 1.366 |
| | 2.327 | 2.598 | 2.327 |
| q=-12e | 0.835 | 1.173 | 0.833 |
| | 1.324 | 2.242 | 1.321 |
| | 2.383 | 2.503 | 2.380 |

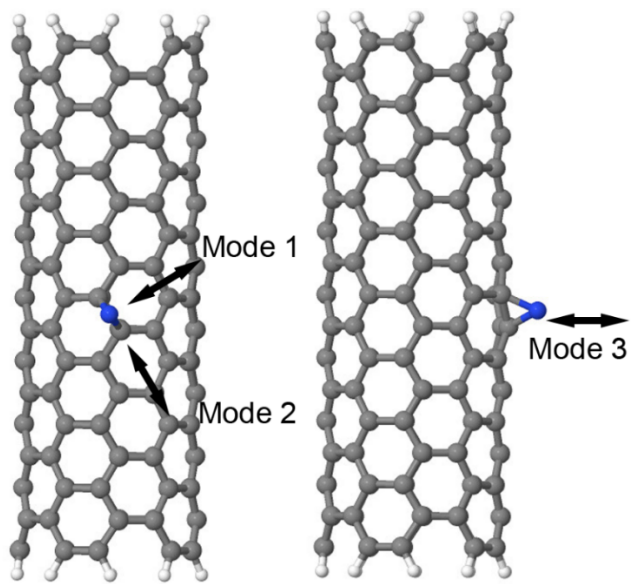

Figure S3 Illustration of three vibrational normal modes of adatom at (5,5) SWCNT

## S4: Transition rates for KMC simulation

**Table S4** Transition rates of adatoms on (5,5) SWCNT at different charges

| Transition type | Transition rates (ns$^{-1}$) | | | | | | |
|---|---|---|---|---|---|---|---|
| | q=0 | q=-2e | q=-4e | q=-6e | q=-8e | q=-10e | q=-12e |
| **1,3->2** | 2.582E-01 | 2.881E-01 | 1.065E+00 | 1.928E-02 | 1.698E-02 | 1.750E-02 | 1.441E-02 |
| **2->1,3** | 8.053E-04 | 2.274E-03 | 3.358E-03 | 6.323E-04 | 2.085E-03 | 6.028E-03 | 1.973E-03 |
| **1->3,3->1** | 6.805E+02 | 4.967E+03 | 4.038E+03 | 4.864E+02 | 3.793E+02 | 2.118E+02 | 3.536E+02 |
| **1->d** | 1.094E-02 | 3.270E-03 | 3.938E-03 | 2.247E-05 | 9.532E-07 | 6.234E-08 | 5.965E-08 |
| **2->d** | 3.067E-05 | 2.581E-05 | 1.242E-05 | 7.371E-07 | 1.170E-07 | 2.147E-08 | 8.168E-09 |
| **3->d** | 1.096E-02 | 3.261E-03 | 3.929E-03 | 2.245E-05 | 9.528E-07 | 6.226E-08 | 5.965E-08 |

**S5: Mulliken charge on adatom with different charges on (5,5) SWCNT**

**Table S5** Charge on adatom with different charges on (5,5) SWCNT

| Adsorption site | Mulliken charge on adatom (e) | | | | | | |
|---|---|---|---|---|---|---|---|
| | q=0 | q=-2e | q=-4e | q=-6e | q=-8e | q=-10e | q=-12e |
| 1 | 0.2567 | 0.1200 | 0.0789 | -0.1942 | -0.2391 | -0.2831 | -0.3165 |
| 2 | 0.2602 | 0.2296 | 0.0456 | -0.0572 | -0.1214 | -0.2113 | -0.2340 |